\documentstyle[aps]{revtex}
\begin{document}
\draft
\input{epsf}
\title{Observing Spontaneous Strong Parity Violation in Heavy-Ion Collisions}

\author{
A. Chikanian                  \unskip,$^{(3)}$
L.E. Finch                    \unskip,$^{(3)}$
R.S. Longacre                 \unskip,$^{(1)}$
J. Sandweiss                  \unskip,$^{(3)}$
J.H. Thomas                   \unskip,$^{(2)}$}
\address{  $^{(1)}$ Brookhaven National Laboratory, Upton, 
New York 11973 \break
  $^{(2)}$ Lawrence Berkeley National Laboratory, 
Berkeley, California 94720 \break
  $^{(3)}$ Yale University, New Haven, Connecticut 06520 \break
}

\date{\today}
\maketitle
\begin{abstract}
We discuss the problem of observing spontaneous 
parity and CP violation in collision systems.  We discuss
and propose observables which may be used in heavy-ion 
collisions to observe such violations, as well as event-by-event
methods to analyze the data.  Finally, we discuss simple monte-carlo
models of these CP violating effects which we have used to develop
our techniques and from which we derive rough estimates of sensitivities
to signals which may be seen at RHIC. 
\end{abstract}


\section{Introduction}
\label{sec-intro}

Parity and CP symmetries have been a source of fascination to physicists
since the discoveries that neither is absolutely 
respected by nature.  \cite{leeandwu,CPviol}.
To date, the violation of these symmetries has only been observed in 
weak interactions, never in strong or electromagnetic interactions.    
Why the CP symmetry should normally be respected by the strong interaction, 
as seems to be the case experimentally, is an outstanding
mystery (the so-called strong CP problem \cite{peccei}), but a topic 
separate from this paper.  
           
Kharzeev et. al. \cite{KharPRL} have proposed that even if the 
strong interactions normally respect CP symmetry (or in the language of the 
field, even if $\bar{\theta}_{QCD}$ \cite{peccei} is equal to zero),
that in hot hadronic matter, metastable states may be formed in which parity 
and CP are spontaneously violated.  

The authors obtained this result using a non-linear sigma model as an effective
Lagrangian, explicitly including a term which incorporates the breaking of the 
axial U(1) symmetry which is present in Quantum Chromodynamics.  They have 
found that within the context of this model, under certain assumptions
about the nature of deconfining and chiral phase transitions, 
metastable states may form which behave as regions in which 
$\bar{\theta}_{QCD}$ is nonzero, 
and so spontaneously break CP symmetry.     
They then further proposed that such regions may be produced in 
relativistic collisions of heavy nuclei.  (The possibility of CP violating 
vacuum states formed in heavy ion collisions is not tied to this mechanism 
and in fact the general idea was raised several years prior \cite{schmidt}.) 

In Reference \cite{KharPRL} and subsequent papers 
\cite{KharPionic,GyulRBRC,buck,ahren,buck2}, methods for observing CP 
violating states in heavy ion 
collisions have been suggested which can be grouped into two categories based 
upon the proposed experimental signatures.  The first group of signatures have 
to do with the observation of parity violating strong decays which are normally 
forbidden and with the observation of anomalous particle production 
\cite{buck,ahren,buck2}.  The second group have to do with direct 
observation of global parity and CP odd observables which may be formed out of 
the momenta of final state particles.  It is with this second group of 
signatures that we are concerned in this paper.

Two mechanisms have been proposed by which strong CP violation can lead 
to observable effects in global variables formed from the momentum vectors of 
the final state particles.  The first is that one of these metastable regions 
which varies in time may give rise to a CP violating asymmetry in $\pi^{+}$ and
$\pi^{-}$ momenta.  This is discussed in Reference \cite{KharPionic} as 
arising through the 'Witten-Wess-Zumino Term' \cite{witten,wess-zumino} and 
led to the proposal for the event-by-event observable   
\begin{equation}
J = \frac{1}{N_{pairs} }\sum_{pairs} 
[ \hat{p}_{+} \times  \hat{p}_{-} ] \cdot \hat{z} 
\label{eq:Jk}
\end{equation}
where the sum is over all possible pairs and each pair contains one observed 
$\pi^{+}$ and one observed $\pi^{-}$.  $\vec{p}_{\pm}$ represents the momentum 
vector of an observed $\pi^{\pm}$ and $\hat{p}_{\pm}$ a unit vector in
the same direction.  $\hat{z}$ must be some other polar 
vector which can be defined in the collision system.
(This particular observable will be discussed further in Section~\ref{sec-obs}.)

Additionally, because a nonzero value for $\bar{\theta}_{QCD}$ 
effectively implies a nonzero value
for $\vec{B}_{c} \cdot \vec{E}_{c}$ (where these are chromo-magnetic and 
chromo-electric fields) \cite{KharPionic,peccei}, it has been proposed 
\cite{GyulRBRC} that this CP asymmetry would be manifested in CP-violating 
deflections of the trajectories 
of quarks which pass through these regions.  This then should ultimately 
appear in CP-odd observables formed out of the momentum vectors of final state 
baryons rather than pions (because pions have equal fragmentation probabilities
from light quarks and antiquarks).   

\subsection{Spontaneous Parity Violation}

Let us end this introduction with a few comments concerning the 
observation of symmetry violation in collision
systems in general (specifically, what type of violations may or may not be 
demonstrated by the observations of non-zero value in a P or CP odd
observable) and the added complications that are related specifically 
to spontaneous symmetry violation.   

In a collision system consisting of two identical spin-0 nuclei, the initial 
state is an eigenstate of parity (in the center-of-mass system), which 
implies that no parity odd observable 
can have a nonzero expectation value with respect to this initial state.  If 
the interaction respects parity symmetry, this must also be true of the final 
state; only if the interaction does not respect parity, producing a final 
state of mixed parity, can the final state show a non-zero expectation value 
for any parity-odd observable.

This conclusion is also true even if the collision system consists of two 
non-identical (spin-0) particles.  Even assuming a nonzero impact parameter, 
only one pseudovector can be defined in the initial state, and it is (by its 
definition) orthogonal to the two available polar vectors.  Therefore if we can
form any pseudoscalar observable in the final state which has a nonzero 
expectation value, this demonstrates a violation of parity in the 
interaction.

If it is true that an expected signal due to parity violation is 
too small to be observed in a single event, as it is here,
then one must clearly look for a cumulative effect over many events.  If the 
interaction violated parity in the same manner for every event (for example if 
$\vec{B}_{c} \cdot \vec{E}_{c}$ were for every event greater than zero rather 
that simply nonzero), then the effect would accumulate over many events and a 
nonzero mean for the distribution of some event-by-event parity odd operator 
could be observed as proof of parity violation.  
If however, as is the case here, the effect is spontaneous and changes 
handedness randomly from event to event with equal probability to be left or
right handed, then the average value over many 
events for a parity odd observable must be zero.  In this case the only 
signatures of the violation we can observe are in the widths of 
event-by-event distributions of P-odd observables.  More discussion of this
point and the methods by which the observation of this change in 
a distribution's width can best be accomplished is found in 
Section~\ref{sec-methods}.

And finally there are two important notes concerning what symmetry violations 
can actually be demonstrated in this collision system.  Clearly, the initial 
state consisting of two nuclei can not be an eigenstate of charge conjugation 
and so our initial state can not have definite CP properties.  Therefore we can
not demonstrate CP violation by observing final states of mixed CP properties 
by the same argument as given above for parity.  However, the nature of the 
expected CP violation can and should lead us in our choice of parity odd 
observables.  Time reversal violation (related of course to CP violation by the
CPT theorem) could in principle be determined by constructing final state 
observables odd under motion reversal but this is complicated by T-odd final 
state phase shifts of the outgoing particles' 
momenta\cite{sachs}.

The second point to note is that if the initial state nuclei are not spin zero,
then the initial state in a given event is not a parity eigenstate and there 
generally will be nonzero pseudoscalar observables which can be constructed 
from the initial state.  For unpolarized beams, this can in principle lead 
to a nonzero expectation 
value for a parity odd observable in each event which may change the 
widths of these event-by-event distributions and be mistaken 
for the signature of spontaneous symmetry violation.  We will come back to 
this point again later with specific observables and a range of observed final 
state momenta in mind, but of course this point is most effectively dealt with 
by colliding beams of spinless nuclei such as Pb(208).   
 
\section{Observables-Old and New}
\label{sec-obs}

As defined in Equation~\ref{eq:Jk} and suggested in reference \cite{KharPRL}, 
the observable $J$ is problematic.  For $J$ to be parity odd, $\hat{z}$ 
must be a real (polar) vector and so must correspond to some uniquely 
definable direction.  The initial suggestion of the $z$ direction as the beam
axis is useful only for collisions of non-identical particles for which
this can be uniquely defined.  

This problem can be rectified as suggested in \cite{KharPionic} by replacing
$\hat{z}$ with a real vector which measures either the particle flow 
($\vec{k}_{+}$) or the charge separation ($\vec{k}_{-}$) of a given event.  
These vectors can be defined using the final state particles' 
momentum vectors as 
\begin{equation}
\vec{k}_{\pm} = \frac{1}{N_{+}}\sum_{+} \hat{p}_{+} \pm 
          \frac{1}{N_{-}}\sum_{-} \hat{p}_{-}
\label{eq:Kpm}
\end{equation}
where the factors $N_{\pm}$ are simply the numbers of observed $\pi^{\pm}$ and 
the sums run over all observed pions of a given species in an event. (We will 
speak about pions in this section for definiteness but as discussed in 
Section~\ref{sec-intro} we can also define and have studied the same 
observables for protons and antiprotons).  Making this replacement of 
$\vec{k}_{-}$ for $\hat{z}$, we have an experimentally definable operator which
is odd under parity, time reversal and CP.  

\begin{figure}
\centering\leavevmode\epsfbox{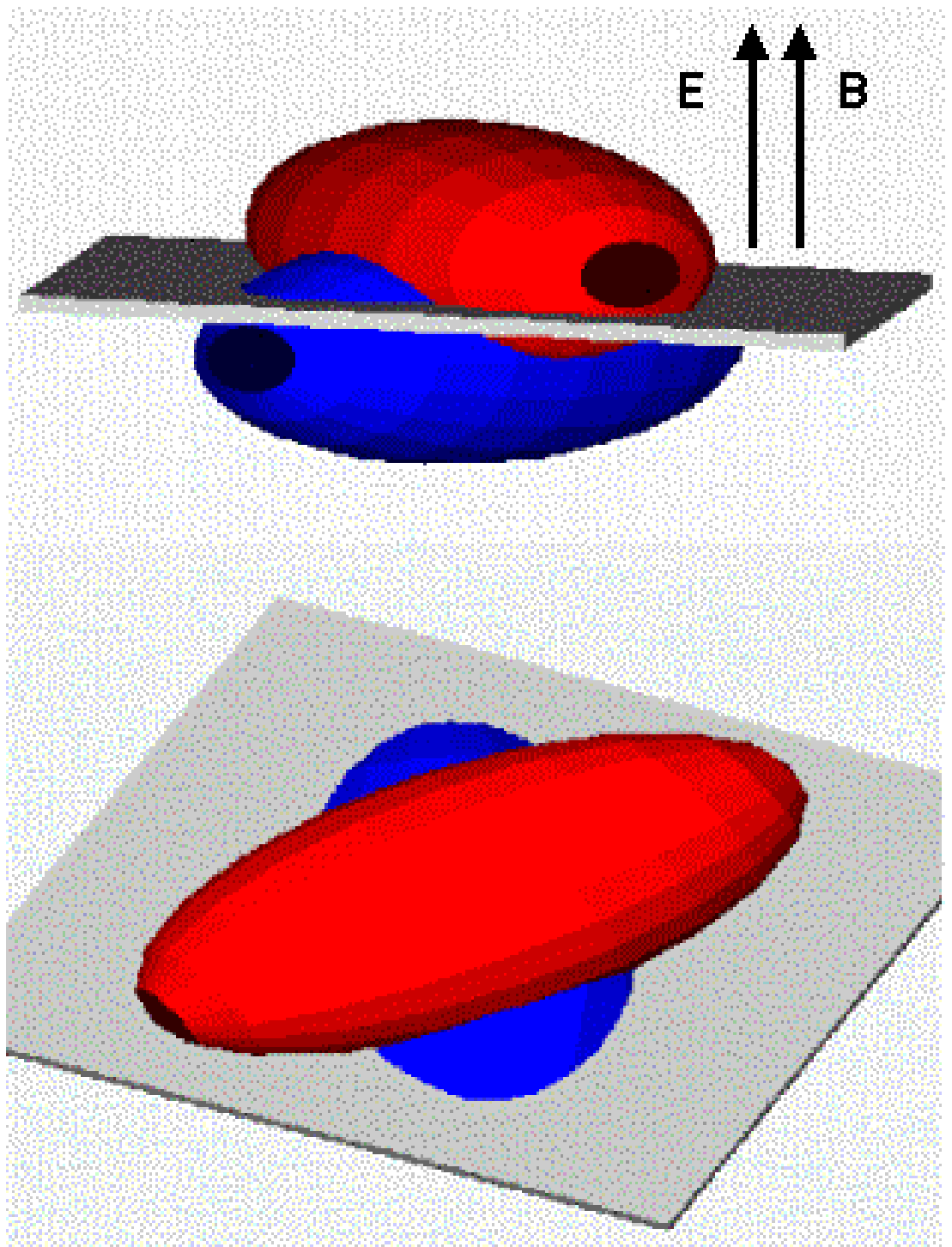}
\caption{Cartoon depicting the momentum space effect of aligned
chromoelectric and chromomagnetic fields on the momentum 
space distributions of quarks and antiquarks.  We imagine that without
the fields present, the two distributions can be pictured as 
overlaping ellipsoids which are elongated along the $p_{z}$-axis.  
The fields then alter these distributions to a state roughly like the
one depicted in a very exaggerated manner in these two cartoons (the two
cartoons show different views of the same situation).  
The plane drawn is the
$p_{x}$-$p_{z}$ plane (assuming the fields are in the y-direction) and is 
included only to help to clarify the cartoon.}
\label{fig:bologna}
\end{figure}

However, for identical particle collisions, the event sum  
$\sum_{pairs} (\hat{p}_{+} \times \hat{p}_{-}) $ results in a vector which 
has an additional fault which can be demonstrated as
follows:  Let the distribution of a given species of particles
emitted from an event be called $g(\theta,\phi)$ where theta is the polar angle
with respect to the beam axis and $\phi$ is the azimuthal angle.  
Then for central collisions on average $g(\theta,\phi)$ is azimuthally
symmetric and $g(\theta)$ = $g(\theta + \pi)$.  By modelling the momentum
kicks that particles receive from the CP violating fields as angular rotations
in momentum space with opposite charged
species being rotated in opposite directions 
(See Figure~\ref{fig:bologna} and related discussion later in this 
section), the contributions to $J$ from various
parts of the $\pi^{+},\pi^{-}$ momentum space exactly cancel one another 
so that the total sum over all
of momentum space is zero \cite{chikandsand}.  
For any given event, these symmetries of $g(\theta,\phi)$ 
will be only approximately realized due to finite statistics (and directed
flow interferes for non-central collisions), but to a large degree this effect 
still persists so that the observable $J$ is forced to be nearly zero by these 
approximate symmetries.  This leads to $J$ as an observable being relatively 
insensitive to the effects of CP violation, and we have noted this effect in 
the results of our simulation models.   

An observable which is similar to $J$ but which is not forced to 
zero by these symmetry constraints
is a tensor observable previously proposed in reference \cite{GyulRBRC} 
defined as
\begin{equation}
T_{ij} =  \frac{1}{N_{pairs}} \sum_{pairs} 
[ \hat{p}_{+} \times \hat{p}_{-} ] \cdot {\hat{n}_{i}}
[ \hat{p}_{+}    -   \hat{p}_{-} ] \cdot {\hat{n}_{j}}
\label{eq:Tzz}
\end{equation}
Here $\hat{n}_{j}$ refers to a unit vector in the $j$th direction.  
$T_{ij}$ is manifestly P and CP odd, since the cross product 
yields an axial vector and the difference of two momentum vectors is a polar 
vector (the unit vectors are axial vectors).  The diagonal components
of this tensor are sensitive to the sorts of symmetry violation which
$J$ was constructed to see;  here we will only discuss $T_{zz}$.  
Each term
in the sum which comprises $T_{zz}$ contributes a value which is roughly 
speaking a measure of the correlation between the longitudinal momentum 
difference of the pair and the azimuthal angle difference of the pair.

If the sum in Equation~\ref{eq:Tzz} is interpreted to be a sum over all possible
pair combinations in an event, then the sum for each component of the tensor 
may be rewritten as a combination of single particle sums rather than a sum 
over pairs.  For example,
\begin{equation}
%
T_{zz} = 
\langle \hat{p}_{x}\hat{p}_{z} \rangle_{+} \langle \hat{p}_{y} \rangle_{-} +
\langle \hat{p}_{x}\hat{p}_{z} \rangle_{-} \langle \hat{p}_{y} \rangle_{+} -
\langle \hat{p}_{y}\hat{p}_{z} \rangle_{+} \langle \hat{p}_{x} \rangle_{-} -
\langle \hat{p}_{y}\hat{p}_{z} \rangle_{-} \langle \hat{p}_{x} \rangle_{+} 
\label{eq:Tzzsingles}
\end{equation}
where for example $\langle  p_{x}p_{z} \rangle_{+} $ is short hand for
$\frac{1}{N_{+}}\sum_{+}(\hat{p}_{+}\cdot\hat{x})(\hat{p}_{+}\cdot\hat{z})$. 
This form of $T_{zz}$ has 
a clear computational advantage for experiments which
may have thousands of particles, and therefore millions of possible
pairs, in a given event. 

$T_{zz}$ also has a very natural interpretation when used, as it was originally
intended, to observe momentum space asymmetries among baryons.  If we consider 
the CP violation as being caused by aligned chromoelectric and chromomagnetic 
fields, we can visualize the effects of these fields in momentum space on the 
distributions of quarks and antiquarks (or, after hadronization, protons and 
antiprotons).  
In the absence of CP violation we imagine that the two different species fill
approximately the same ellipsoid in momentum space, elongated along the beam
($z$) axis.  With the addition of these CP violating fields, however, the 
ellipsoids are (roughly speaking) displaced to a form such as that shown in 
Figure~\ref{fig:bologna}.  The 
chromoelectric field moves the two species apart along the direction of the 
field, and the chromomagnetic field rotates the two species' momentum space 
distributions in opposite directions around the field axis (note that
if we assume that the momentum distribution of particles in the bubble
were isotropic then no effect would be observable). 
It is this combined action of the 'lift' along the field axis 
with the 'twist' around the field axis which $T_{zz}$ is sensitive to. 

This picture also leads to the idea that equation~\ref{eq:Tzzsingles} may be 
rewritten in a coordinate system defined by the charge flow axis
($\hat{k}_{-}$), and then it may be beneficial to selectively keep only those 
terms from Equation~\ref{eq:Tzzsingles} which lay along this axis.  This 
process yields a different observable, 
\begin{equation}
K_{twist} =   
\frac{1}{N_{+}}\sum_{+}( p_{y}k_{-,x} - p_{x}k_{-,y}) -
\frac{1}{N_{-}}\sum_{-}( p_{y}k_{-,x} - p_{x}k_{-,y})
\label{eq:Ktwist}
\end{equation}
where the vector $\vec{k}_{-}$ is defined by Equation~\ref{eq:Kpm}.

We can also slightly alter the construction of $J$ and so create a different
pseudoscalar observable which is not forced to zero by symmetry considerations.
We simply introduce into each term of the sum which defines $J$ 
a factor which depends on the relative momentum space position of the 
the two pions involved.  Specifically, an extra factor of $-1$ is introduced 
depending on whether the two pion momenta are in the same or opposite 
directions with respect to the beam (z) axis in the collision center-of-mass 
frame.  Following this prescription we can then define the vector $\vec{J_{c}}$
as
\begin{equation}
\vec{J}_{c} = \frac{1}{N_{pairs}} \sum_{pairs} [(\hat{p}_{+} \times 
\hat{p}_{-}) sgn(p_{z,+}  p_{z,-})] 
\label{eq:Jc}
\end{equation}
and we note that the scalar $\vec{J}_{c} \cdot \vec{k}_{-}$ is a P and
CP odd observable which we propose as another alternative to  $J$, intended to 
capture the same physics without its practical difficulties. 

And finally, we have also studied the observable proposed in reference
\cite{Voloshin} defined as 
\begin{equation}
V_{+}  =   
( \langle \vec{p}_{t} \rangle_{+,y>y_{cm}} \times 
\langle \vec{p}_{t} \rangle_{+,y<y_{cm}} ) \cdot \hat{z} 
\label{eq:V}
\end{equation} 
(Here, $\hat{z}$ is a unit vector along the z axis which is chosen arbitrarily
along one of the beam directions, and so is an axial vector.)
This observable has the advantage that it can be defined for a single
particle species (as written, it only involves $\pi^{+}$) 
and so avoids assumptions about how
particles and antiparticles will be affected differently by the
fields.  A disadvantage is the resulting assumption that the 
particles with rapidity greater than $y_{cm}$ will be deflected 
oppositely to those travelling in the opposite direction
(tantamount to assuming that the CP violating region will generally 
sit at $y = y_{cm}$).

We have found from our model simulations (see Sections~\ref{sec-models} 
and~\ref{sec-results}) that under most circumstances these observables
behave quite similarly and generally do not differ from one another in 
sensitivity by more than a factor of a few in any given situation, with none 
of them being systematically better than the others.  It will likely be useful
to study experimental data with all of them, as they do provide useful cross
checks on one another and of course a factor of a few in statistics is not
to be taken lightly. 
We do not discuss here other proposed observables such
those defined in \cite{KharPionic} because we have not studied these in any 
detail. 

\section{Simulation Models}
\label{sec-models}

In order to develop these observables as well as to make rough estimates
of the sensitivity that we may expect a RHIC experiment to have
to these effects,
we have developed some simple models to simulate the effects of the
proposed strong CP violation.  These models are not intended to
be realistic descriptions of the collision and CP violating dynamics, but
rather to lead to similar final state asymmetries which we can use to
estimate observable signals.

We base Model $I$ on the idea that the CP violating region will exist as
a 'bubble' inside the collision region which contains a nonzero value of 
$\vec{E} \cdot \vec{B}$.  These fields then affect the
trajectories of quarks or hadrons which pass through this bubble.  
We model the color fields as electromagnetic fields and so let them
provide forces by acting on the electric charges of the particles; thus
mimicking the net effect that the strong fields would have on the particles'
trajectories.  Modelling the fields in this manner is an assumption 
but we believe not an unreasonable one in 
light of the discussion in Section~\ref{sec-intro}
concerning the predicted net effects that the color forces will have on the
pion and baryon momentum spectra.

The initial distribution of particles in momentum space 
for Model $I$ is taken from the event
generator HIJING \cite{hijing}.  Initial positions for these particles 
are then picked randomly from a uniform distribution
over a spherical collision volume (note than no 
position-momentum correlations are included in this simple model).  The CP 
violating bubble is then included as a static, randomly positioned 
spherical sub-region of this collision volume in which there are 
nonzero aligned electric and magnetic fields. 
As charged particles move from their initial positions, if 
they happen to cross through this CP violating bubble their trajectories are 
altered by the presence of the aligned fields. 
The field strength and size of these regions is chosen to correspond to
theoretical predictions about these CP violating regions and their color 
fields; they are determined as described in Section~\ref{sec-results}. 

Additionally, to explore the idea advocated in \cite{GyulRBRC} of examining 
the trajectories of final state baryons rather than pions, the model is 
altered in the following way for further studies:  The initial proton and 
antiproton distributions are again taken from HIJING.  Each baryon is 
decomposed into its constituent quarks which are then spread through the 
collision region.  After each quark is allowed to traverse the CP violating 
bubble, if appropriate given its initial position and trajectory, the quarks 
are re-coalesced into a baryon, which as a result of this process receives a 
net momentum kick equal to the sum of the kicks given to the constituent 
quarks.     

To summarize then, we model the fields as producing opposite impulses on 
oppositely charged particles, producing CP violating asymmetries in both 
pionic and proton-antiproton
observables.  (Probably the mechanism is somewhat closer to the truth for 
$p/\bar{p}$ where the proposed color fields really should act 
oppositely on the different species.  For pions, we may simply view this 
mechanism as a way to 
produce the expected asymmetries caused by CP violation). 

Models $II$ and $III$ are similarly based on 
the idea of embedding into events metastable 
bubbles which have a vacuum where CP is violated.  For these models, we use 
RHIC::EVENT\cite{rhicevent} (an improved version of HIJET\cite{hijet}) as the 
basic event generator. 

This generator was then customized for CP violation studies.  In each generated 
event, the particles within a certain spatial region are 'tagged' to be 
turned into a bubble with CP violating vacuum (for details 
see\cite{rhicevent}) which has the momentum and baryon numbers of the 
particles it replaces.  This bubble consists of quarks uniformly 
distributed in a region of aligned color electric and magnetic fields (of 
randomly chosen direction) that alter the quarks' momentum similarly to what 
was described for pions in model $I$ (oppositely for quarks and antiquarks).  
When a quark reaches the boundary of the bubble, it forms a pion which retains 
the effect of the impulse imparted to the quark from the color fields (again, 
for details see\cite{rhicevent}).  For these models, the 
expected net flow of pion 
charge \cite{KharPionic} is obtained by 
allowing the momentum impulses given to quarks to affect positive pions and the
impulses given to anti-quarks to affect only negative pions.  

Using this underlying model, we have studied the effects of 
different assumptions concerning the
momentum distributions of pions emitted from the bubble.  Two assumptions that
were studied were a spherically symmetric Boltzmann distribution with a 
temperature of 70 MeV (this is Model $II$), and a Landau 
expansion with a transverse temperature of 150 MeV and a distribution along 
the beam axis which is flat over four units of rapidity (Model $III$).

For Model $II$, if the color fields are of uniform strength throughout 
the bubble, no signal due to the fields' effects may be seen on the final 
momentum of the pions  (imagine Figure~\ref{fig:bologna} with initially 
spherically symmetric distributions in momentum space, 
so that the rotation caused by the color 
magnetic field could not be detected.).  In this case we have broken this 
symmetry by making the field strengths constant only in the transverse 
direction and letting them have a strong longitudinal dependence.  For 
Model $III$, as in Model $I$, the necessary asymmetry is built in to the pion 
momentum distribution, so that such a field dependence need not 
be assumed.

To obtain quantitative results from these models concerning sensitivities
at RHIC, we assume an acceptance of -1 to +1 in rapidity and .120 to 
1 GeV/c in transverse momentum, similar to that of STAR \cite{star}. 

\section{Experimental Methods}
\label{sec-methods}

As described in Section~\ref{sec-intro}, the spontaneous nature of this proposed
CP violation makes it impossible to see a nonzero expectation value in the 
distribution of any CP odd observable which will accumulate from event to 
event.  This, combined with the fact that we expect any single event 
to have far too small a signal to be statistically significant (this is
clearly confirmed by our model simulations even under extreme assumptions
about the strength of the CP violation), means we have 
to turn to event-by-event techniques to look for a signal.  

\begin{figure}
\centering\leavevmode\epsfbox{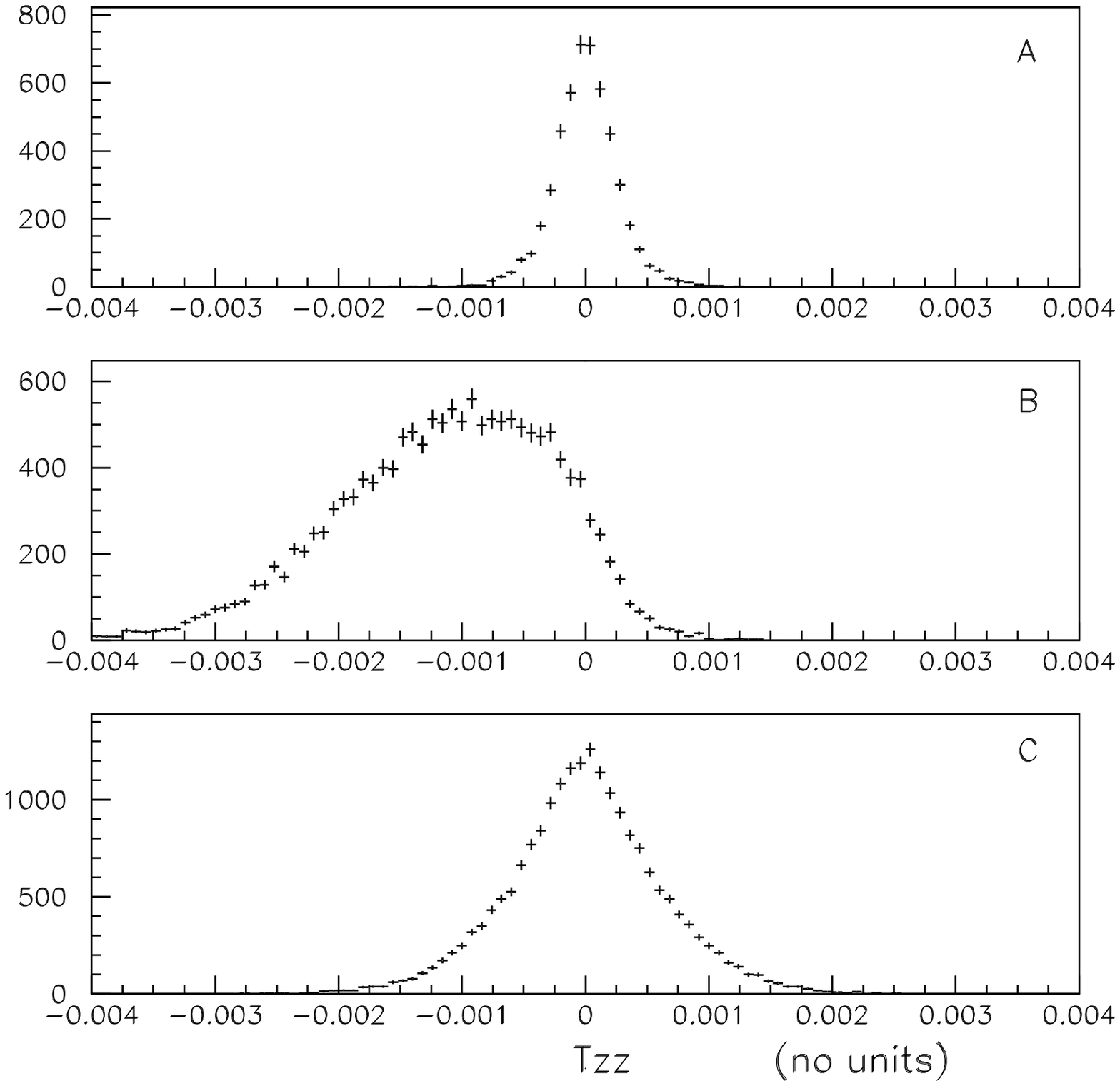}
\caption{Illustration of the problem of observing parity violation when
in each event the violation is equally likely to be right-handed or
left-handed.  Panel A shows the distribution of $T_{zz}$ for simulated events
with no parity violating fields.  Panel B shows the same distribution with
$\vec{B} \cdot \vec{E} > 0 $ for every event.  For panel C, $\vec{B} 
\cdot \vec{E} $ is as likely greater than as less than zero in each event; 
this is expected to be the case if strong CP violation occurs.  The difference
between panels A and C then is simply in the width of the distribution, and
this is the basic signal of parity violation (shown here in exaggeration) 
for which we are searching with the methods described here.}  
\label{fig:widths}
\end{figure}

Because any signal of spontaneous CP violation will be contained 
essentially in the width of the event-by-event
distributions of CP odd observables (as illustrated is 
Figure~\ref{fig:widths} ) such as those described above, 
the most straightforward
method is to measure the width of one of these distributions and compare it to
a reference 'no-signal' distribution.  Of course, then the problem becomes one
of how to create a bias free reference distribution with which we can compare.  

\subsection{Mixed Event Reference Distributions}
\label{sec-mixed}

Perhaps the easiest way to create a reference distribution is 
to use real events to create a pool of "mixed-events".  These
are fake events which are constructed from small pieces (ideally, 
just one track) taken from many different
real reconstructed events.  In this manner one creates a pool of mixed events 
which have the global features of real events but absent any of the
correlations between tracks, including those caused by CP violation. 
Thus in the presence of CP violation, the distribution of values
of any of our observables should be wider in real events than it is
in mixed events.

The presence or absence of a signal for parity violation can be determined 
quantitatively in this method then by simply forming histograms of the 
distribution of (for example) $T_{zz}$ for real events and for mixed events 
and then taking the difference of these two histograms.  This 
difference histogram can then be compared channel by channel with a value of 0 
and a $\chi^{2}$ value obtained for the difference.  If this process were 
repeated many times on signal-free independent samples of equal size, the 
values of $\chi^{2}$ should form a gaussian distribution about the number
of degrees of freedom ($N_{dof}$) in the fit with a width given by 
$\sqrt(2N_{dof})$  (this is assuming a sufficiently large number of non-zero 
bins).  For any single data sample then, the value 
$(\chi_{2} - N_{dof})/ \sqrt(2N_{dof})$ gives a quantitative measure of the 
number of standard deviations away from zero signal. This measure does
depend somewhat on the binning chosen for histograms, but the variations
for any reasonable choice of binning tends to be well less than a factor
of 2.

In our simulation models, using this method with any of the observables 
described above shows by far the largest signal
in the presence of CP violating bubbles of any method we have tried, but
this is quite misleading:  Any effect which makes the momentum space structure
different in real events than in mixed events can in principle lead
to a signal in this method.  This certainly includes practical experimental
effects which will be discussed in Section~\ref{sec-reality}.  
However, this also includes other physics effects
such as the presence of jets and fluctuations of mean $p_{T}$ in real events.
Additionally, a positive signal using any of the observables described above 
can be produced by the 
presence of a nonzero $\vec{B}$ field without a correlated $\vec{E}$, or vice 
versa.  (This signal is due to a change in the width of the distributions 
that comes about not because a slightly 
nonzero mean in each event is added on top of fluctuations as would
be the case for correlated $\vec{B}$ and $\vec{E}$, but simply 
because the presence of a single field causes the size of the 
event-by-event fluctuations about zero to increase) ; 
one field alone does not imply a CP violation 
but does produce a signal in this mixed event method. 

This final point is somewhat subtle so we will attempt to clarify it, using
$\vec{J}_{c} \cdot \vec{k}^{-}$ for definiteness.  If the real event 
distribution of $|\vec{k}_{-}|$ has an rms width larger than the same 
distribution from mixed events (which it does in our simulations if we turn on
only an $\vec{E}$ field since this is essentially what $k^{-}$ measures), 
this by itself will give a larger width to the distribution of 
$\vec{J}_{c} \cdot \vec{k}^{-}$ in real compared with mixed events, 
without any event-by-event correlation in the directions of 
the two vectors in real events.  We could in principle avoid this problem by 
switching to an observable such as $\hat{J}_{c} \cdot \hat{k}^{-}$ which 
measures only the angular correlation between the two vectors, but this is 
more effectively done by changing the way in which the reference distributions 
are made to the method described in Section~\ref{sec-oth}

We conclude then that using the mixed event method as here described 
might be useful to set an upper limit in the absence of a signal
(if some non-trivial practical problems can be overcome), but
the observation of a signal using this technique could never by itself be 
taken as strong evidence for parity violation.  

\subsection{Subevents}
\label{sec-sub}

A second method, which for our purposes is much more robust, 
is another standard event-by-event technique generally referred 
to as the Subevent Method \cite{voloshinsubs?}.  The method is quite simple: we
in some manner parse the tracks from an event into two 
subevents and calculate the value of one of our observables, say $K_{twist}$, 
for each subevent.  To look for a signal then, we look at 
whether there is a significant covariance between the distributions of 
$(K_{twist})_{sub1}$ and $(K_{twist})_{sub2}$.  Equivalently, we can observe 
the distribution of event-by-event values of 
$(K_{twist})_{sub1} \cdot (K_{twist})_{sub2} $.  If the mean of this 
distribution is significantly shifted away from zero, a parity violation is 
implied.  

The idea behind this method then is this:  the expectation value of 
$(K_{twist})$ in any event or subevent in which parity is not violated will be
zero (and the distribution of values should be symmetric about zero).  
So if we choose two random uncorrelated subevents from an event and calculate 
$K_{twist}$ for each of these, we randomly sample two 
such distributions, and 
$(K_{twist})_{sub1} \cdot (K_{twist})_{sub2}$ should then also be 
symmetrically distributed about zero.  If ,however, there is CP violation in 
an event, we expect both subevents to yield values for $(K_{twist})$ which 
are on average shifted slightly in the same direction, yielding a distribution 
for $(K_{twist})_{sub1} \cdot (K_{twist})_{sub2}$ which, given enough events, 
will have a mean significantly greater than zero.  

Different methods of choosing subevents have their own
advantages and disadvantages as discussed briefly in \cite{voloshinsubs?}.
Two possible variations are (i) Choosing the subevents randomly from the 
available tracks, and thus forming two subevents which overlap in momentum 
space. (ii) Dividing an event into subevents by partitioning momentum space, 
ideally with some gap in momentum space between the two.  
Because the parity violation may be localized in momentum space, variation (i) 
may have an advantage in sensitivity in real data.  Variation (ii) however, 
has a clear advantage in avoiding correlations between the two subevents, and 
for a study such as this where a false signal is clearly highly undesirable 
this is an important advantage.  Also, (ii) may be easily generalized to a 
larger number of smaller subevents per event which in principle is useful for 
looking for parity violating correlations over a momentum range smaller than 
an entire event.  

The subevent method avoids most of the pitfalls of the mixed event method 
discussed above, but the price is that it takes significantly more
events to produce an effect in the subevent method. (Just to
clarify again: the underlying 
reason for this is that the subevent method in our
simulations is sensitive only to the presence of aligned color $E$ and $B$ 
fields which implies CP violation; the mixed event method is sensitive not 
only to this, but also to the presence of only one of these fields, which does 
not by itself imply CP violation.)
This will be discussed more quantitatively in Section~\ref{sec-results}.
                                                              
\subsection{Other Reference Distributions}
\label{sec-oth}
 
There are other methods for producing reference distributions which avoid 
the problems of the mixed event method as described above.  For example, when 
using a pseudoscalar observable which measures the event-by-event 
correlation of two vectors (we'll use 
$\vec{J}_{c}\cdot \vec{k}_{-}$ as an example), one can form the reference 
distribution by using the vector $\vec{J}_{c}$ from a given event and the 
vector $\vec{k}_{-}$ from a different event; there are several similar 
variations of this method which retain this basic principle.  
We find this technique to be similar to the subevent method in its
sensitivity to CP violation, and it is similarly not sensitive to
the false signals that the mixed event method is.


\section{Simluation Results and Expected Sensitivities}
\label{sec-results}

For our studies with model $I$, the size of the CP violating region and the 
strength of the fields in this region can be varied for a systematic 
study of the signal strength as a function of these parameters, and some
of the results from these studies are shown in  Table~\ref{tab:tab1}.  
Clearly, there is considerable uncertainty in choosing reasonable values
for these parameters.  The nominal value 
for the field strength we have taken as suggested by D. Kharzeev
\cite{KharPrivate}; both $E$ and $B$ fields are strong 
enough to give a 30 
MeV/c kick in momentum 
to a relativistic quark which traverses the length of the region (for
simplicity, from now
on we will write field strengths in units where this value is equal to one).  
For the bubble size, our nominal choice is to use a bubble radius equal to 
1/5 or 2/5 of the radius of the 'collision region'.  These sizes are chosen 
so that the fields affect about 100 pions in an event, as suggested in 
\cite{KharPionic}.
We list in Table ~\ref{tab:tab1} the approximate number of 
RHIC Au+Au events that were required in our simulations to produce an effect
from CP violation at the $3\sigma$ level in the observable $T_{zz}$.  We also 
list results for other larger choices of bubble size and field 
strength mainly to demonstrate the scaling of the number of required
events versus these parameters.  In the subevent case, we find that
the strength of the signal (as measured by shift of the distribution mean
away from zero) scales as the field strength to the fourth power, as expected;
the value of these observables is affected linearly by the strength of
both the $E$ and $B$ fields and the shift of the subevent distribution
scales as the square of the event-wise distribution.  
Recall that as discussed in Section~\ref{sec-methods}, although the mixed-event 
method is clearly the more sensitive, a signal in this method could not be 
taken as strong evidence for parity violation.  

Although we show only
the results for $T_{zz}$ in this table, the
results for $K_{twist}$, $V$, and $\vec{J}_{c}\cdot \vec{k}_{-}$ are 
roughly consistent with these; variations on the order of factors
of a few between the observables in some cases are observed.  

In Table~\ref{tab:tab2} we show a comparison between the predictions of the 
three different models with comparable choices of field strength and bubble 
sizes.  The nominal values used for models $II$ and $III$ were  
established in a similar manner to those for model $I$, with the bubble size 
chosen again so that approximately 100 pions are produced by quarks exiting the 
bubble.  For the comparison shown here we have used a bubble radius of 1/5 in 
Model $I$ and the actual observable compared is $K_{twist}$.  
This comparison gives some idea of the variations we can obtain even from
somewhat similar models:  model $II$ (non-uniform field strength in
the longitudinal direction) predicts considerably better 
sensitivity than Model $I$, while Model $III$ (pions 
with a Landau distribution) gives us results more similar to model $I$.  

We see from Tables~\ref{tab:tab1} and~\ref{tab:tab2} that even 
with the mixed event method, we would not
expect to see any signs of the presence of CP violation with less than a 
few$\times 10^{5}$ central RHIC events.  For strong evidence of parity 
violation (meaning a signal with the subevent method or a using a more
robust reference distribution method) the number of events needed seems
to be more 
like a minimum of a few $\times 10^{7}$ 
central RHIC events.   

Some comments about these results are in order.  Firstly, we emphasize again 
that the transition from the theory advocated in \cite{KharPRL} to these models
requires several assumptions and if these are changed the results could be 
quite different.  For example, we can also choose for nominal values of the 
bubble size and strength those values which would produce a signal at the 
$10^{-3}$ level in the observable $J_{c}$ for pion pairs coming from the bubble 
(as suggested in \cite{KharPionic}) as 
opposed to the $10^{-2}$ level which is produced by model $I$ using the
parameters from the top row of Table~\ref{tab:tab1};  
this would clearly lead to a considerably more pessimistic outlook for
the number of events needed.  


Finally, we should mention our results from using model $I$ to
model the effects on protons and antiprotons as described in 
Section~\ref{sec-models}.  Clearly, the main drawback in using baryons
rather than pions is the small relative number produced in the collision.
In the case of bubble radius = 2/5 and field strength of 1
(the more optimistic 'nominal' case from above) this model predicts
that we could observe a $3\sigma$ signal using the mixed event method in 
approximately 3 million central events (compared to about 300K events with 
pions).  We conclude then that we have little chance of observing a CP 
violation signal for 
protons and antiprotons, particularly using any of the more robust methods.  

\section{False Signals}
\label{sec-reality}

Of course a very important point in studies such as these is the possibility of
something other than actual parity violation faking a positive signal.

\subsection{Experimental Inefficiencies}

Experimental inefficiencies may be divided into 
two categories: Single track 
inefficiencies 
and correlated inefficiencies.

We have attempted to simulate various patterns of 
single track inefficiencies as a function of polar and azimuthal angle as well
as differing inefficiencies for different charge species and have been unable
to produce a fake signal of any sort from these effects alone.
Furthermore, even if a pattern exists which can cause a fake signal, provided 
that it is static in time, it clearly should not behave as a signal which 
changes handedness event-to-event and therefore should be observable as a 
shift of the mean of the event-by-event distributions of our CP odd variables 
rather than just a broadening of these distributions.  
Considering these factors, we conclude
that single track inefficiencies alone are not a significant concern for 
providing a false signal which is indistinguishable from a true parity 
violation (the same arguments also apply to single track measurement errors).

More concerning are correlated efficiencies, and specifically track merging and 
splitting (the results of track recognition software mistaking two nearby
tracks for a single track or mistaking a single track for two similar tracks). 
Either of these can generally cause a fake signal for 
the mixed event method as can in principle any effect that makes the momentum 
space distribution of tracks different in real events compared with mixed 
events.  This can to some extent be compensated for when forming mixed events, 
but this is a paramount practical concern about using the mixed event method. 

In principle track splitting may also cause difficulties with the subevent 
method; clearly if one of the two versions of a split track is assigned to 
each of two subevents, the subevents will contain a correlation that may lead 
to correlations in their values for parity odd observables.  This was mentioned
in Section~\ref{sec-sub} as motivation for forming subevents which are 
separated from one another in momentum space; provided that both tracks 
resulting from a split track are contained within the same subevent, track 
splitting should not cause a false signal for the subevent method which we
can not distinguish from a real signal.

\subsection{Physics concerns}

With a cartoon similar to that shown in Figure~\ref{fig:bologna}, we can 
visualize in momentum space the combined effects of directed flow together
with a charge 
separation which increases the radial momentum of one charge species 
while decreasing the other.  The combined result of these effects and finite 
detector acceptance can mimic the 'twist' shown in Figure~\ref{fig:bologna},
but not the 'lift'.  
This generally would still not be confused with a CP violating effect because 
for this confusion to occur there must be some additional effect which 
distinguishes 'up' from 
'down' along the twist axis.  The differentiation could 
in principle be provided by 
asymmetric acceptances and efficiencies but these effects should be 
identifiable (for example, creating mixed events only out of events with nearly
common reaction planes and processing these events by the subevent method
should also show a signal if the signal is caused by these effects but not if 
it is a true CP violation).  And ultimately, any false effect which has flow as 
one of its root causes should be distinguishable by its 
known dependence on collision centrality \cite{starflow}.  

It is also true that many hyperons will be produced in a typical central event 
and these do in fact undergo parity violating weak decays.   However, this 
process alone should 
not be able to lead to a parity violating effect in global event observables 
unless some facet of hyperon production or polarization depends on production 
angle in a way that itself violates parity.  

\subsection{Nuclei with Spin}

If the two colliding nuclei are not spin zero, then it is no longer true that 
we can form no pseudoscalars in the initial state, and so it no longer follows 
that the expectation value of every pseudoscalar observable in any given 
collision is zero, but (provided the beams are not polarized), the average 
value over many events must still be zero.  This, then, is a signal that may 
in principle appear in the 'width' of the distribution of some pseudoscalar 
observable but not as a shifted mean value; exactly the type of effect we are 
concerned with observing.  This is clearly a relevant point because the 
colliding beams at RHIC are composed of Au(197), a spin-(1/2) nucleus.   

We should then consider what effect these initial state spins might have on the
observables with which we are concerned here.  For the observable $V$ defined 
in Equation~\ref{eq:V}, a left-right asymmetry in the production of $\pi^{+}$ 
and $\pi^{-}$ as is known to occur in polarized p-p collisions \cite{adams} 
is one mechanism by which nonzero spin could lead to an effect which would 
mimic CP violation.  For example (labelling the two beam directions in 
a collision 'forward' and 'backward'), if in a given collision forward going
$\pi^{+}$ moved preferentially to the right with respect to the beam 
while backward moving $\pi^{+}$ moved preferentially down, this would lead to 
nonzero value for $V$ calculated from the $\pi^{+}$.  

We can very roughly 
estimate the size of this effect using the 
available data \cite{adams} and the resulting phenomenological descriptions 
\cite{anselmino} to extrapolate to the momentum space region relevant for 
these studies.  The asymmetry is characterized for a proton beam with spin 
measured to be 'up' or 'down' by
\begin{equation}
A_{N} = - \frac{1}{cos \phi} 
\frac{N_{\uparrow}(\phi) - N_{\downarrow}(\phi)} {N_{\uparrow}(\phi) + 
N_{\downarrow}(\phi)}\label{eq:An}
\end{equation} 
where $\phi$ is the angle between 'up' and the pion production plane and 
$N_{\uparrow}(\phi)$ is the number of pions produced at $\phi$ when
the proton is measured to have spin up.
This asymmetry can be quite large in p-p collisions; typically, 
$A_{N} \approx $ 0.3
for $x_{F} \gtrsim 0.5 $ and $p_{T} \gtrsim 700 MeV/c$.  However, for typical 
values of $x_{F}$ ($\lesssim$ 0.05) at $p_{T}$ ($\lesssim$ 1 GeV/c) that are 
appropriate here, the asymmetry is much smaller; 
we can take a conservative value of the asymmetry for 
$\pi^{+}$ as $10^{-2}$.  

Using this value of $10^{-2}$ for $A_{N}$ of the $\pi^{+}$ and assuming 
the beams to be unpolarized, we can with our 
simulation models estimate the number of events needed to see a $3\sigma$ 
signal in $V_{+}$ due to this effect:
In each event we randomly choose polarization vectors for the forward
and backward beams.  We then alter the event so that forward (backward)
travelling $\pi^{+}$ have an asymmetry of $A_{N} = 0.01$ with respect to the 
polarization direction of the forward (backward) travelling beam.
We find that assuming an asymmetry of this size 
will give a $3\sigma$ signal for the 
subevent method in roughly $10^{9}$ events. 
Furthermore, this result is under the assumption that 
spin (1/2), A=200 nuclei will 
produce as large an asymmetry as a protons, which seems an 
extremely conservative assumption. 

We conclude therefore that it is extremely unlikely that this particular 
mechanism presents a practical difficulty with the amount of data likely to be 
collected by RHIC experiments.  We suspect that the this may be generally true 
of any CP-mimicking effects which may be caused
by spin, because the small size originates chiefly in the small 
expected value for 
spin-induced asymmetries in the central rapidity region which will be observed 
at RHIC.  Clearly, though, if a potential signal for strong parity 
violation is observed in a collision of spin (1/2) nuclei, this point should be
addressed more thoroughly.  

\section{Summary}

We have discussed methods of identifying spontaneous strong parity 
violation in heavy-ion collisions and argued that this 
violation may in principle be uniquely identified by standard event-by-event 
techniques for any colliding system of spin-0 nuclei.  
We have discussed and introduced observables which should be useful
in looking for such a violation.  In our simulations, under 
certain assumptions about the strength of the violation, we find that for
the mechanism proposed by Kharzeev et. al., it is possible that we would
see a signal which we would consider as strong evidence of parity violation
with a few times $10^{7}$ central events in a detector such 
as STAR, although some of the assumptions which lead to this number may be
optimistic.  At any rate, this is an tremendously interesting effect to
be searched for and should be studied with as much RHIC data as becomes
available.  

We have investigated likely sources of false signals and have thus far found
none which we expect could not be distinguished from a true parity violation,
save in principle effects due to the spin of the incoming nuclei (and
we believe that in practice this will also not be a significant problem). 
\label{sec-summary}


\section{Acknowledgements}
This work was supported in part by grants from the Department of Energy 
(DOE) High Energy Physics Division and the DOE nuclear division.

%

\newpage


\begin{table}
\caption{Estimated number of central RHIC events needed to see a $3\sigma$ 
signal for the observable $T_{zz}$, as a function of 'bubble' size and field 
strength in Model $I$
(assuming a STAR-like acceptance).  The top two cases are those that we 
discuss in the text as the 'nominal' cases for these parameters.
A value for the top case in the Subevent method is not included because
a sufficient number of simulation events have not been run. }
\label{tab:tab1}
\begin{tabular}{|c|c|c|c|}
Field Strength& $R_{dom}/R_{tot}$ &Events Needed (Mixed Event)  &
		Events Needed(Subevent) \\ \tableline \tableline
1   &  1/5  &  10M  &         \\ \tableline
1   &  2/5  &  50K  &  30M    \\ \tableline
1   &  3/5  &   3K  & 350K    \\ \tableline
2   &  3/5  & 300   &   1.5K  \\ \tableline
3   &  3/5  & 100   & 200     \\ 
\end{tabular}
\end{table}

\begin{table}
\caption{Results from the three different simulation models for the
number of central RHIC events needed to observe a $3\sigma$ signal resulting
from CP violation using the mixed-event
method and 'nominal' values for bubble size and field strength as 
described in the text.  These results are assuming a detector with
acceptance similar to that of STAR.}
\label{tab:tab2}
\begin{tabular}{|c|c|}
Model             &  Number of Events Needed  \\ \tableline \tableline
$I$               &  10M \\ \tableline
$II$              &  200K \\ \tableline
$III$             &  2M   \\
\end{tabular}
\end{table}

\end{document}